\newif\ifdraft
\newif\ifpreprint
\newif\iffull
\preprinttrue 
\fulltrue  

\ifpreprint
\documentclass[ALICE,manyauthors]{cernphprep}
\else
\documentclass[final,3p,12pt]{elsarticle}
\fi
\usepackage{hyperref}

\usepackage{lineno} 
\usepackage{rotating}
\usepackage{color}
\usepackage{graphicx}

\ifpreprint
\newcommand{\pip}          {$\pi^{+}$}
\newcommand{\pim}          {$\pi^{-}$}

\newcommand{\lam}          {$\Lambda$}

\newcommand{\vzero}        {V$^0$}
\newcommand{\csi}          {$\Xi$}
\newcommand{\csim}         {$\Xi^{-}$}
\newcommand{\csip}         {$\overline{\Xi}^{+}$}

\newcommand{\om}           {$\Omega$}
\newcommand{\omm}          {$\Omega^{-}$}
\newcommand{\omp}          {$\overline{\Omega}^{+}$}

\newcommand{\ssbar}        {$s\overline{s}$}
\newcommand{\ptopion}      {$\mathrm{p}/\pi$}
\newcommand{\xitopion}     {$\Xi/\pi$}
\newcommand{\omtopion}     {$\Omega/\pi$}

\newcommand{\s}            {$\sqrt{s}$}
\newcommand{\pt}           {$p_{\rm T}$}
\newcommand{\GeV}          {GeV/$c$}
\newcommand{\MeV}          {MeV/$c^2$}
\newcommand{\dedx}         {d$E$/d$x$}
\newcommand{\dndy}         {d$N$/d$y$}

\newcommand{\pp}           {pp}
\newcommand{\pPb}          {p--Pb}
\newcommand{\pBe}          {p--Be}

\newcommand{\AAcoll}       {\mbox{A--A}}
\newcommand{\PbPb}         {\mbox{Pb--Pb}}
\newcommand{\AuAu}         {\mbox{Au--Au}}

\newcommand{\snn}          {$\sqrt{s_{\mathrm{NN}}}$}
\newcommand{\snnbf}        {\ensuremath{\mathbf{{\sqrt{s_{\mathbf NN}}}}}}

\newcommand{\pseudorap}    {\mbox{$\left | \eta \right | $}}

\newcommand{\Npart}        {$N_\mathrm{part}$ }
\newcommand{\avNpart}      {$\langle N_\mathrm{part} \rangle$}

\else
\input{multistrangePbPbPLBCommands.tex}
\fi
\graphicspath{{./figs/}}

\ifdraft
\fi

\begin{document}%
%
\ifpreprint
\begin{titlepage}
\PHnumber{2013-134}      
\PHdate{July 18, 2013}              
%
\title{Multi-strange baryon production at mid-rapidity \\
in \PbPb\ collisions at \snnbf\ = 2.76 TeV}
\ShortTitle{Multi-strange baryon production in \PbPb\ collisions at \snn\ = 2.76 TeV}   
%
\Collaboration{ALICE Collaboration%
         \thanks{See Appendix~\ref{app:collab} for the list of collaboration members}}
\ShortAuthor{ALICE Collaboration\footnote{See Appendix A for the list of collaboration members}}      
\else
\begin{frontmatter}
\title{Multi-strange baryon production at mid-rapidity \\ in \PbPb\ collisions at \snn\ = 2.76 TeV}
\input{Alice_Authorlist_2013-Jul-12-ELSEVIER.tex}  
\fi
\begin{abstract}
The production of \csim\ and \omm\ baryons and their anti-particles in \PbPb\ collisions at \mbox{\snn\ = 2.76 TeV} has 
been measured using the ALICE detector.
The transverse momentum spectra at mid-rapidity ($|y|$~$<$~0.5) for charged \csi\ and \om\ hyperons 
have been studied in the range 0.6~$<$~\pt~$<$~8.0 \GeV\ and 1.2~$<$~\pt~$<$~7.0 \GeV,
respectively, and in several centrality intervals
(from the most central 0-10\% to the most peripheral 60-80\% collisions). 
These spectra have been compared with the predictions of recent hydrodynamic models. 
In particular, the Krak{\'o}w and EPOS models give a satisfactory description
of the data, with the latter covering a wider \pt\ range. 
Mid-rapidity yields, integrated over \pt, have been determined.
The hyperon-to-pion ratios are similar to those at RHIC: they rise smoothly with centrality
up to \avNpart\ $\sim$ 150 and saturate thereafter. 
The enhancements (yields per participant nucleon relative to those in \pp\ collisions)
increase both with the strangeness content of the baryon and with centrality, but are less pronounced 
than at lower energies.

\ifpreprint
\fi

\end{abstract}

\ifpreprint
\end{titlepage}
\else
\end{frontmatter}
\fi

\ifdraft
  \linenumbers
\fi

\section{Introduction}

The study of strange and multi-strange particle production in relativistic heavy-ion 
collisions is an important tool to investigate the properties of the strongly interacting 
system created in the collision.
Particle spectra provide information both about the temperature of the system and about collective flow. 
In particular they reflect conditions at kinetic freeze-out, i.e. the point in the expansion 
where elastic collisions cease. 
Collective flow is addressed by hydrodynamic models, and depends on the 
internal pressure gradients created in the collision. 
The effects are species-dependent, so new data on multi-strange baryons at LHC energies can bring new constraints to models.

The enhancement of strangeness in heavy-ion collisions was one of the earliest proposed signals for 
the Quark-Gluon Plasma~\cite{Rafelski:1982,Koch:1983,Koch:1986}. It rests on the expectation that in a deconfined
state the abundances of parton species should quickly reach their equilibrium va\-lues, resulting in a higher abundance 
of strangeness per participant than what is seen in proton-proton interactions. In this picture equilibration takes 
place quickly owing to the low excitation energies required to produce $q\bar{q}$ pairs. 
However, it was shown that, at the same entropy-to-baryon ratio, the plasma in equilibrium does not contain more
strangeness than an equilibrated hadron gas at the same temperature~\cite{Redlich:1985,McLerran:1987,Lee:1988}.
Strangeness enhancements have indeed been observed by comparing central heavy-ion collisions with \pBe\ and \pp\ reactions both 
at the SPS~\cite{WA97:1999,Antinori:2006,Antinori:2010,Alt:2005,Alt:2008,Anticic:2009} 
and at RHIC~\cite{Adams:2004,Adams:2007,Abelev:2008}. 
Over the past 15 years, it has been found that the hadron yields in central heavy-ion collisions follow the
expectation for a grand-canonical ensemble~\cite{PBM:2004}, increasingly well as a function of the collision energy, indicative
of a system in equilibrium. At the same time it was understood that, for \pp\ collisions, canonical suppression effects
are important~\cite{Hamieh:2000} and account for the overall hyperon enhancement. 
The progressive removal of these effects also qualitatively describes the increase in strangeness yields with centrality in \PbPb, 
although at RHIC
it was noted that canonical suppression could not successfully reproduce all the features of particle production~\cite{Abelev:2009,Agakishiev:2012}.
At lower energies a better description of the system size dependencies could be achieved using 
a \mbox{core-corona} model~\cite{Becattini:2008,Aichelin:2009,Becattini:2009}.
These pictures can now be re-examined at the much higher LHC energy. 
The most straightforward expectation would be equilibrium values
for the yields of strange particles in central \PbPb\ collisions, combined with reduced canonical suppression in 
proton-proton collisions. 
In this Letter, after an introduction to the ALICE detector and a description of the analysis techniques 
used to identify strange particles via their decay topology, the multi-strange baryon \pt\ spectra are presented. 
Spectra in five different centrality intervals are compared with hydrodynamic models and the corresponding mid-rapidity yields are given.
Their ratios to the interpolated yields for \pp\ interactions at the same centre-of-mass energy, normalized to the number of 
participant nucleons, are used to obtain the enhancement plot as used at lower energies. 
In addition, we study the dependence on centrality of 
the hyperon-to-pion production ratio at mid-rapidity and compare these results with predictions.

\section{The ALICE experiment}

The ALICE experiment was specifically designed to study heavy-ion collisions at the LHC.
The apparatus consists of a central barrel detector, covering the pseudorapidity window \pseudorap\ $<$ 0.9, 
in a large solenoidal magnet providing a 0.5 T field, and a forward dimuon spectrometer with a separate 0.7 T dipole magnet.
Additional forward detectors are used for triggering and centrality selection.
The first LHC heavy-ion run took place at the end of 2010 with colliding Pb ions accelerated to a centre-of-mass energy 
per nucleon of \mbox{\snn\ = 2.76 TeV}. The analysis described in this paper uses data from this first heavy-ion run 
where events in a wide collision centrality range were collected, and is based on 
the information provided by the sub-detectors mentioned below.

Tracking and vertexing are performed using the full tracking system. It consists of the Inner Tracking System
(ITS), which has  six layers of silicon detectors and the Time Projection Chamber (TPC). Three different technologies are used for the ITS:
Silicon Pixel Detectors (SPD), Silicon Drift Detectors (SDD) and Silicon Strip Detectors (SSD).
The two innermost layers (at average radii of 3.9 cm and 7.6 cm, covering \pseudorap $<$ 2 and \pseudorap $<$ 1.4, respectively) 
consist of pixel detectors.
These are used to provide high resolution space points (12 $\mu$m in the plane perpendicular to the beam direction
and 100 $\mu$m along the beam axis). The two intermediate layers consist of silicon drift detectors,
and the two outermost layers of double-sided silicon microstrips. Their radii extend from 15 cm to 43 cm
and they provide both space points for tracking and energy loss for particle identification.
The precise space points provided by the ITS are of great importance in the definition of secondary vertices.
The TPC is a large cylindrical drift detector whose active volume extends radially from 85 cm to 247 cm, and from \mbox{$-$250 cm} to \mbox{+250 cm}
along the beam direction. For a charged particle traversing the TPC, up to 159 space points can be recorded.
These data are used to calculate a particle trajectory in the magnetic field,
and thus determine the track momentum, and also to measure \dedx\ information for particle identification.

The SPD layers and the VZERO detector 
(scintillation hodoscopes placed on \mbox{either} side of the interaction region, covering \mbox{2.8 $<$ $\eta$ $<$ 5.1} 
and \mbox{$-$3.7 $<$ $\eta$ $<$ $-$1.7})
are used for triggering. The trigger selection strategy is described in detail in~\cite{Aamodt:2011}.
In addition, two neutron Zero Degree Calorimeters (ZDC) positioned at $\pm$ 114 m from the interaction point
are used in the offline event selection.
A complete description of the ALICE sub-detectors can be found in~\cite{Aamodt:2008}.

\section{Data samples and cascade reconstruction}

The analysis was performed on the full sample recorded during the 2010 \PbPb\ data taking.
Only events passing the standard selection for minimum bias events were consi\-dered.
This selection is mainly based on VZERO and ZDC timing information to reject beam-induced backgrounds and 
events coming from parasitic beam interactions (``satellite'' collisions). 
The VZERO signal is required to lie in a narrow time window of about 30~ns around the nominal collision time, 
while a cut in the correlation between the sum and the difference of the arrival times in each of the ZDCs 
allows to remove satellite events.
In addition, a minimal energy deposit of about 500 GeV in the ZDCs is required to further suppress 
the background from electromagnetic interactions (for details, see~\cite{Aamodt:2011,Aamodt:2010}).
Only events with a primary vertex position within 10 cm from the centre of the detector along the beam
line were selected; this ensures good rapidity coverage and uniformity for the particle reconstruction
efficiency in the ITS and TPC tracking volume.
In order to study the centrality dependence of multi-strange baryon production, these
events were divided into five centrality classes according to the fraction of the total inelastic
collision cross-section: 0-10\%; 10-20\%; 20-40\%; 40-60\%; 60-80\%.
The definition of the event centrality is based on the sum of the amplitudes measured in the VZERO
detectors, as described in~\cite{Aamodt:2011,PbPb276TeVcentrality}.
The final sample in the 0-80\% centrality range corresponds to approximately 15~$\times$~10$^6$~\PbPb\ collisions
at \snn\ = 2.76~TeV.
For each centrality class the average number of participant nucleons, \avNpart, is calculated from a
Glauber model~\cite{PbPb276TeVcentrality,Bialas:1976,Miller:2007}. This is important for comparisons
since the number of participants is often used as a centrality measure at lower energies or in different collision systems.

\begin{figure}[hbt!f]
\begin{center}
\includegraphics[width=7.8cm]{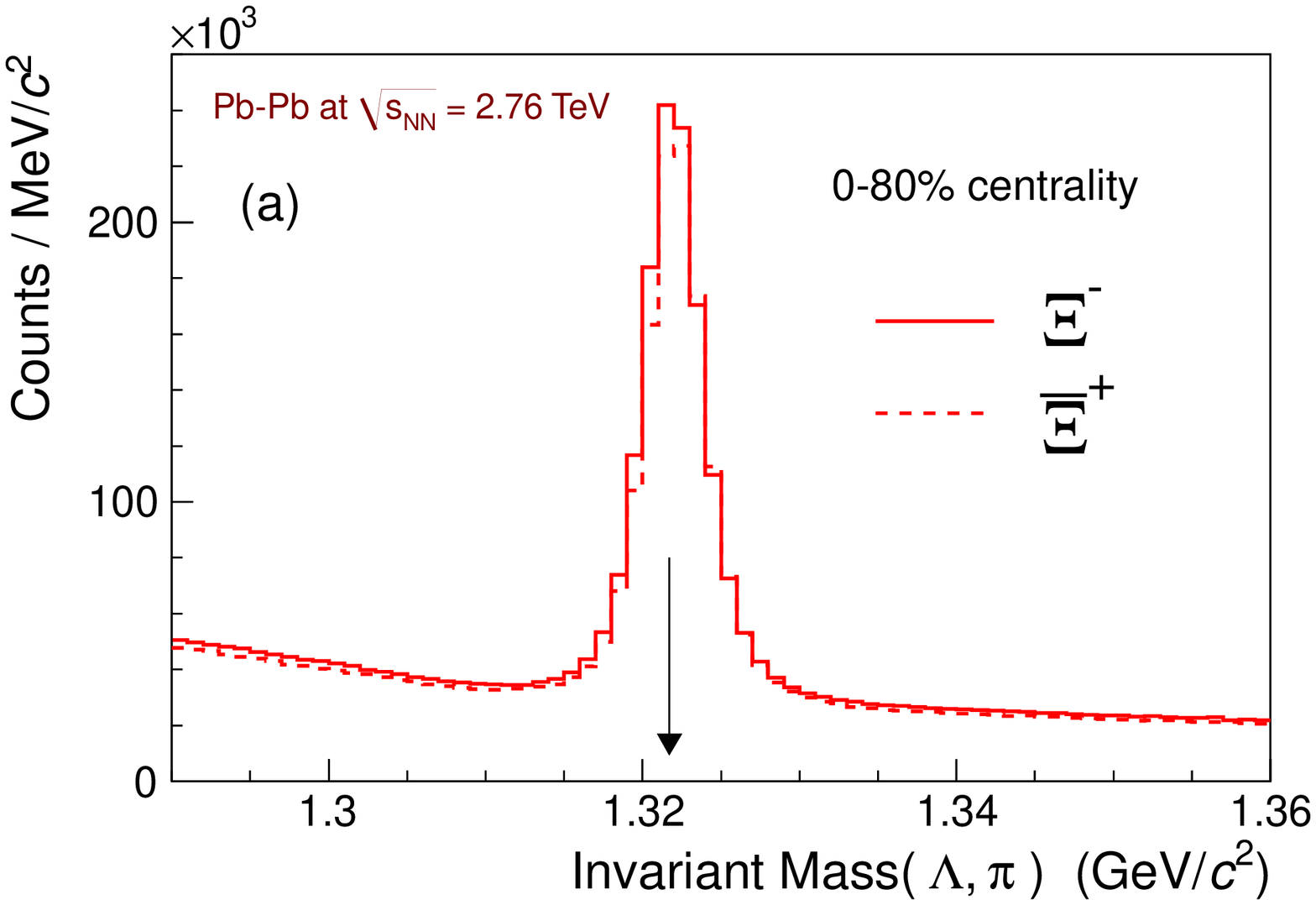}
\includegraphics[width=7.8cm]{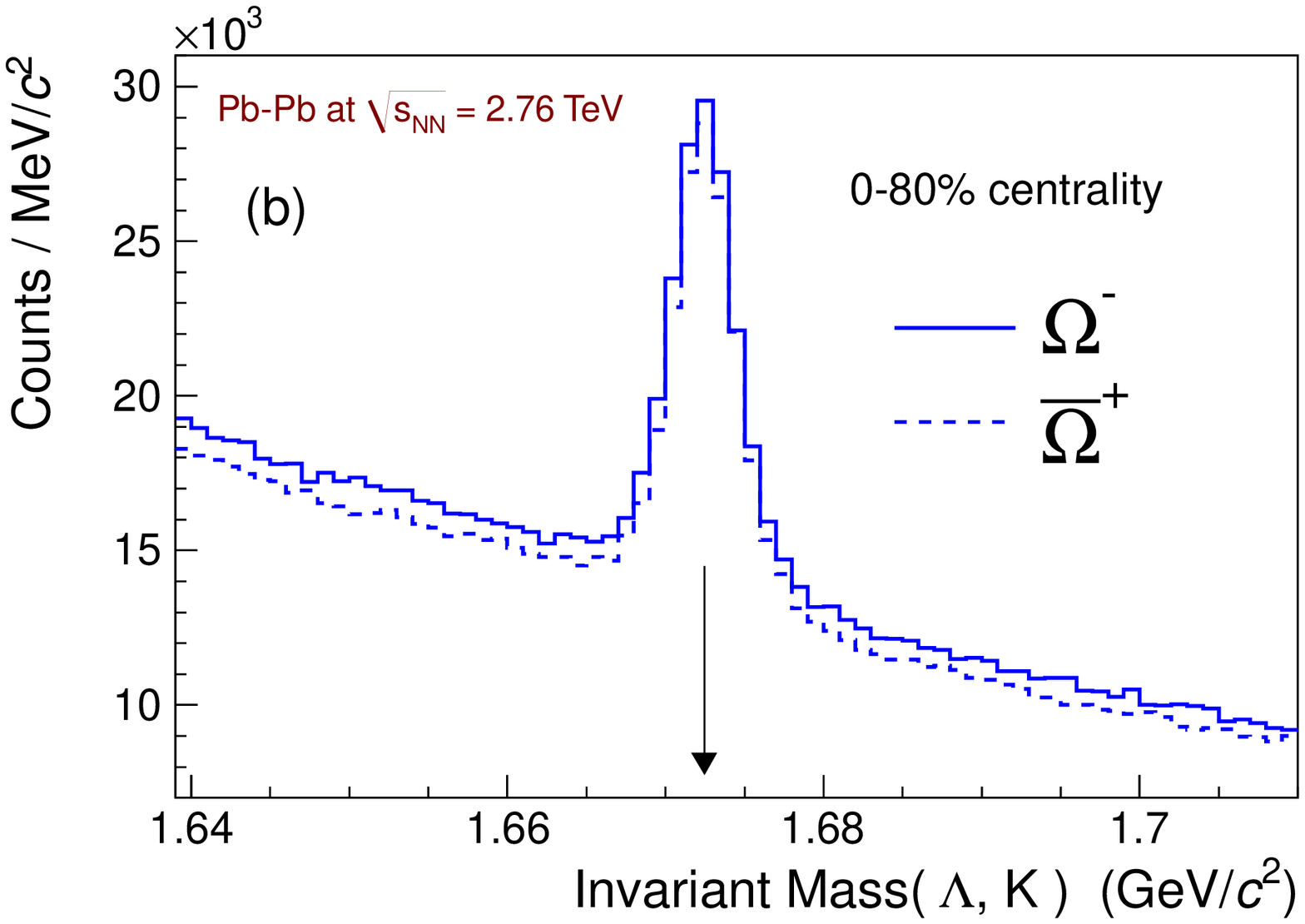}
\caption{\label{Fig_msPbPb_invmass}
Invariant mass distributions for \csi\ (a) and \om\ (b) selected candidates from 0-80\% most central \PbPb\ 
collisions at \mbox{\snn\ = 2.76 TeV}. The plots are for candidates in the rapidity interval $|y|$~$<$~0.5, 
at \mbox{\pt\ $>$ 0.6} and 1.2 \GeV\ for \csi\ and \om, respectively. The arrows point to the PDG mass values.}
\end{center}
\end{figure}

Multi-strange baryons are measured through the reconstruction of the cascade topology of the following
weak decays into final states with charged particles only:
\mbox{$\Xi^{-} \rightarrow \Lambda + \pi^{-}$} (branching ratio 99.9\%) and
\mbox{$\Omega^{-} \rightarrow \Lambda + \mathrm{K}^{-}$} (67.8\%)
with subsequent decay \mbox{$\Lambda \rightarrow \mathrm{p} + \pi^{-}$} (63.9\%),
and their charge conjugates for the anti-particle decays.
The resulting branching ratios are 63.9\% and 43.3\% for the \csi\ and the \om, respectively.
Candidates are found by combining charged tracks reconstructed in the ITS and TPC volume.
Topological and kinematic restrictions are imposed, first to select the ``\vzero'' (\lam\ candidate V-shaped decay), 
and then to match it with one of the remaining secondary tracks (``bachelor'' candidate).
The distance of closest approach (DCA) between the two \vzero\ daughter tracks, or between the \vzero\ and the bachelor track, 
or the \vzero\ and the primary vertex position,
as well as the \vzero\ and cascade candidate pointing angles (PA) with respect to the primary vertex 
position, are among the most effective selection variables. Pre-defined fiducial windows around the
Particle Data Group (PDG)~\cite{Beringer:2012} mass values are set, both to select the \lam\ in the cascade candidates 
($\pm$~5~\MeV) and to reject \om\ candidates that match the \csi\ hypothesis ($\pm$~8~\MeV).  
In addition, each of the three daughter tracks is checked for compatibility with the pion, kaon or proton 
hypotheses using their energy loss in the TPC.  
The selection procedure, while similar to that utilized for the \pp\ sample~\cite{Abelev:2012},
is optimized for the higher multiplicity environment of the \PbPb\ collisions, which required
tightening the cuts on the DCA and PA variables.
In particular, all the cuts are fine-tuned in the final analysis, and cross-checked with Monte Carlo simulations, 
in order to find the best compromise between the
combinatorial background minimization and the significance of the signals.
The invariant mass distributions of the candidates for all particle species passing the selection cuts
are shown in Fig.~\ref{Fig_msPbPb_invmass}.
The \mbox{signal-to-background} ratio, integrated over $\pm$ 3$\sigma$, is 4.1 for the \csi\ and 1.0 for the \om.
The combinatorial background for anti-particles is approximately 5\% smaller than for particles, 
over the whole measured \pt\ range. This difference has been found to 
increase rapidly when going to the lowest momenta, consistent with the different absorption
cross sections for baryons and anti-baryons within the detector material.

Data are partitioned into the five centrality bins mentioned above and, for each centrality, into different \pt\ intervals.
To extract the raw yields, a symmetric region around the peak \mbox{($\pm$ 3$\sigma$)} is defined by fitting the distribution
with the sum of a Gaussian and a polynomial. The background is determined by sampling the regions on both sides
of the peak; in these regions, whose width and distance from the peak vary with centrality, \pt\ and particle species,
the invariant mass distribution is fitted with a second order polynomial (first order for high \pt\ bins).
The raw yield in each \pt\ and centrality bin is then obtained by subtracting the integral of the background
fit function in the peak region from the total yield in the peak region obtained from bin counting.

\begin{figure}[hbt!f]
\begin{center}
\includegraphics[width=9.5cm]{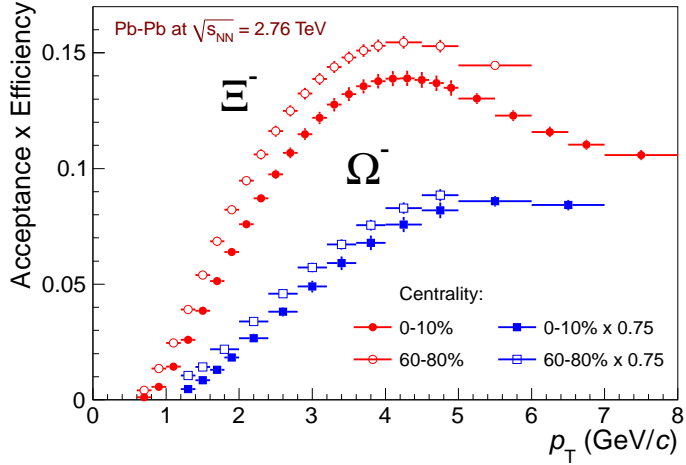}
\caption{\label{Fig_msPbPb_efficiencyvsptcent}
Acceptance $\times$ efficiency factors for \csim\ (circles) and \omm\ (squares) at mid-rapidity as a function of \pt, 
both for the most central 0-10\% (full symbols) and the most peripheral 60-80\% (open symbols) \PbPb\ collisions. 
The points already take into account the branching ratios of the corresponding measured decay channels.
Those for the \omm\ are also scaled by a factor of 0.75, to avoid overlap with the \csim\ at high \pt.}
\end{center}
\end{figure}

A correction factor, which takes into account both the detector acceptance and the reconstruction efficiency
(including the branching ratio of the measured decay channel), is determined for each particle species as a function of \pt, 
and also in different rapidity intervals to verify that the correction varies by less than 10\% with rapidity. 
This is true for $|y|$ $<$ 0.5 for all particles with \pt\ $>$ 1.8 \GeV; for lower transverse momenta, a narrower
rapidity range ($|y|$ $<$ 0.3) has been chosen.
Corrections were determined using about $3\times 10^{6}$ Monte Carlo events, generated using HIJING~\cite{Wang:1991}
with each event being enriched by one hyperon of each species, generated with a flat \pt\ distribution. 
The ``enriched'' events were then processed with the same reconstruction chain used for the data events.
To check that the results are not biased by the presence of such injected signals,
the correction computed with the enriched events and that obtained using a ``pure'' HIJING sample
were compared in the low \pt\ region (below 3 \GeV) 
and found to be compatible. Both samples have then been used to maximize the total available statistics for the computation of the correction.
As an example, Fig.~\ref{Fig_msPbPb_efficiencyvsptcent} shows the resulting \mbox{acceptance $\times$ efficiency} factors as a function 
of \pt\ for \csim\ and \omm, both for the most central (0-10\%) and the most peripheral (60-80\%) classes.
The uncertainties correspond to the total statistics of the Monte Carlo samples used to compute the correction. 
The curves for the anti-particles are compatible with those for particles.
The values are found to decrease with increasing event centrality, as expected.
Compared to the correction applied in the 7 TeV \pp\ collision analysis~\cite{Abelev:2012}, they are smaller by a factor 
between 2.5 and 3 in the most peripheral class of the \PbPb\ sample, basically because of the tighter selection cuts
in the heavy-ion analysis.

\section{Corrected spectra and systematic uncertainties}

\begin{figure}[hbt!f]
\begin{center}
\includegraphics[width=7.8cm]{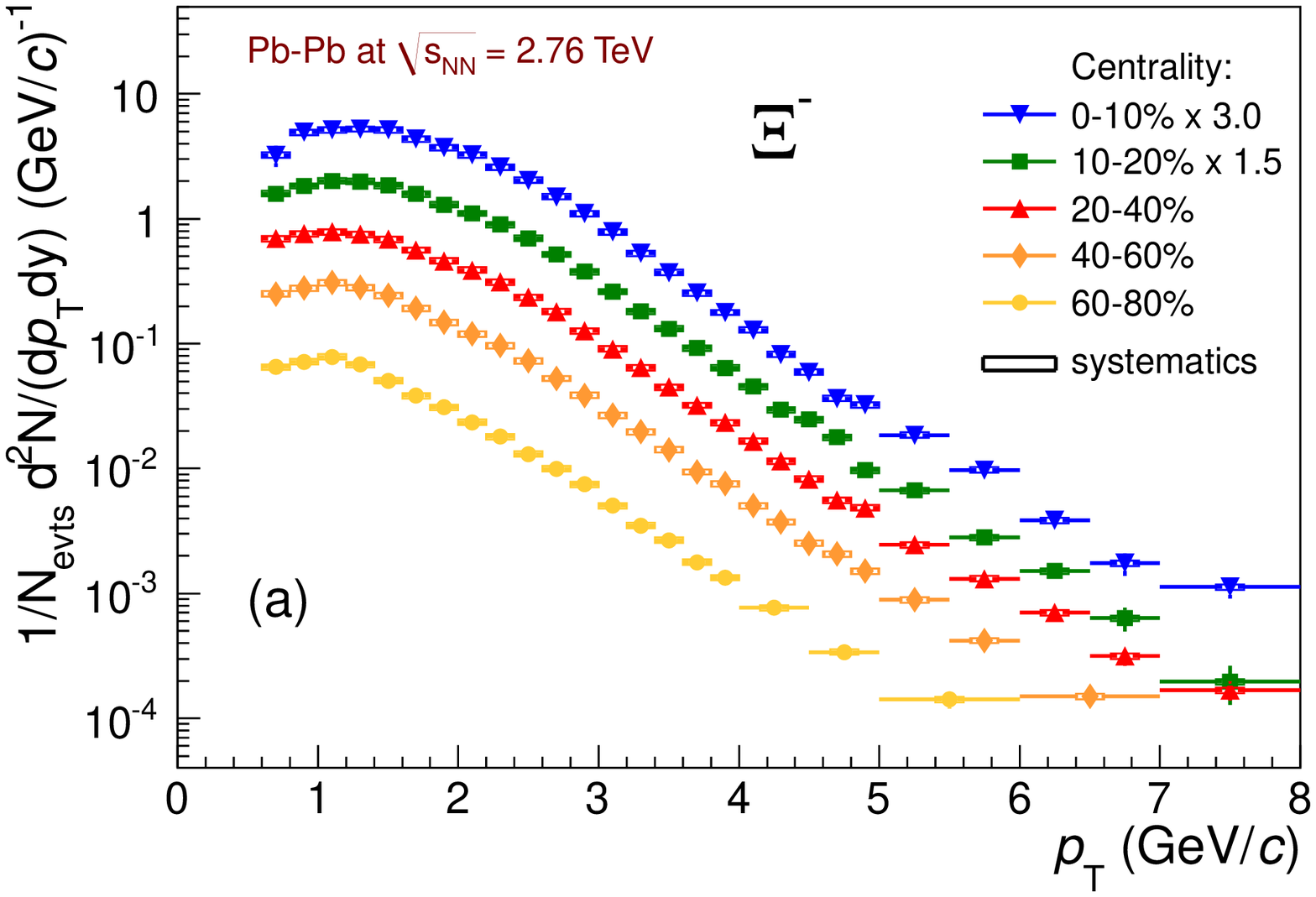}
\includegraphics[width=7.8cm]{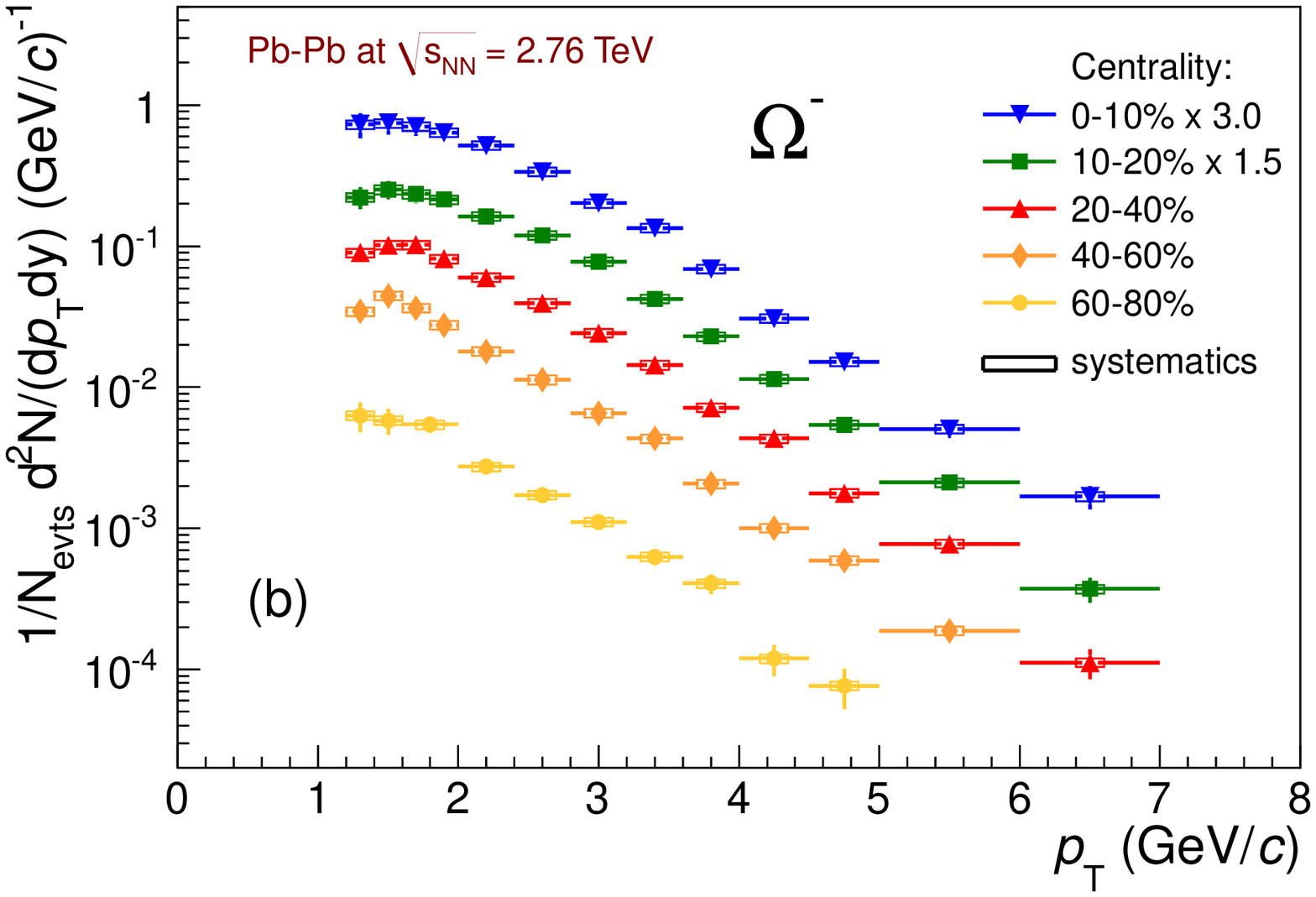}
\includegraphics[width=7.8cm]{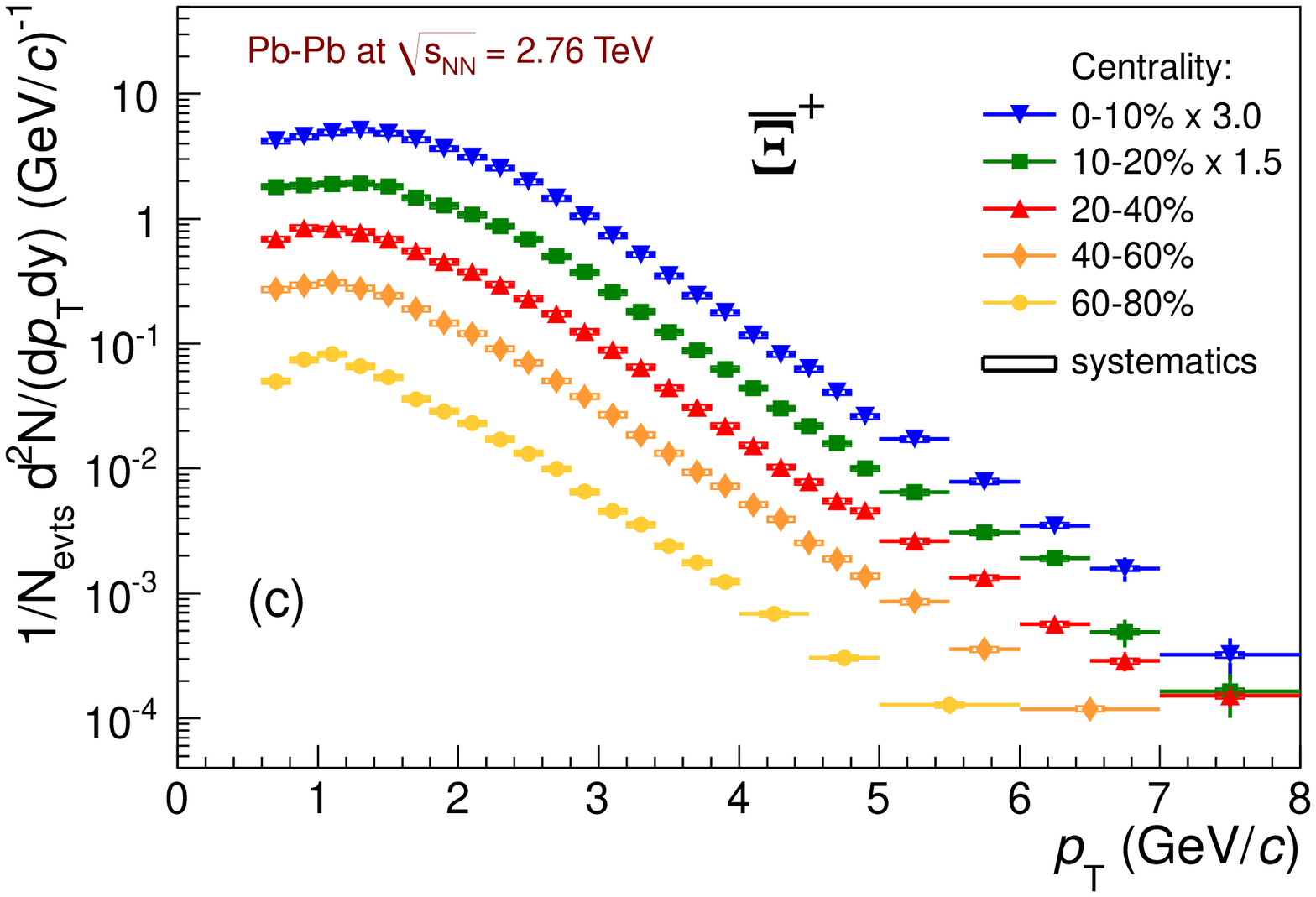}
\includegraphics[width=7.8cm]{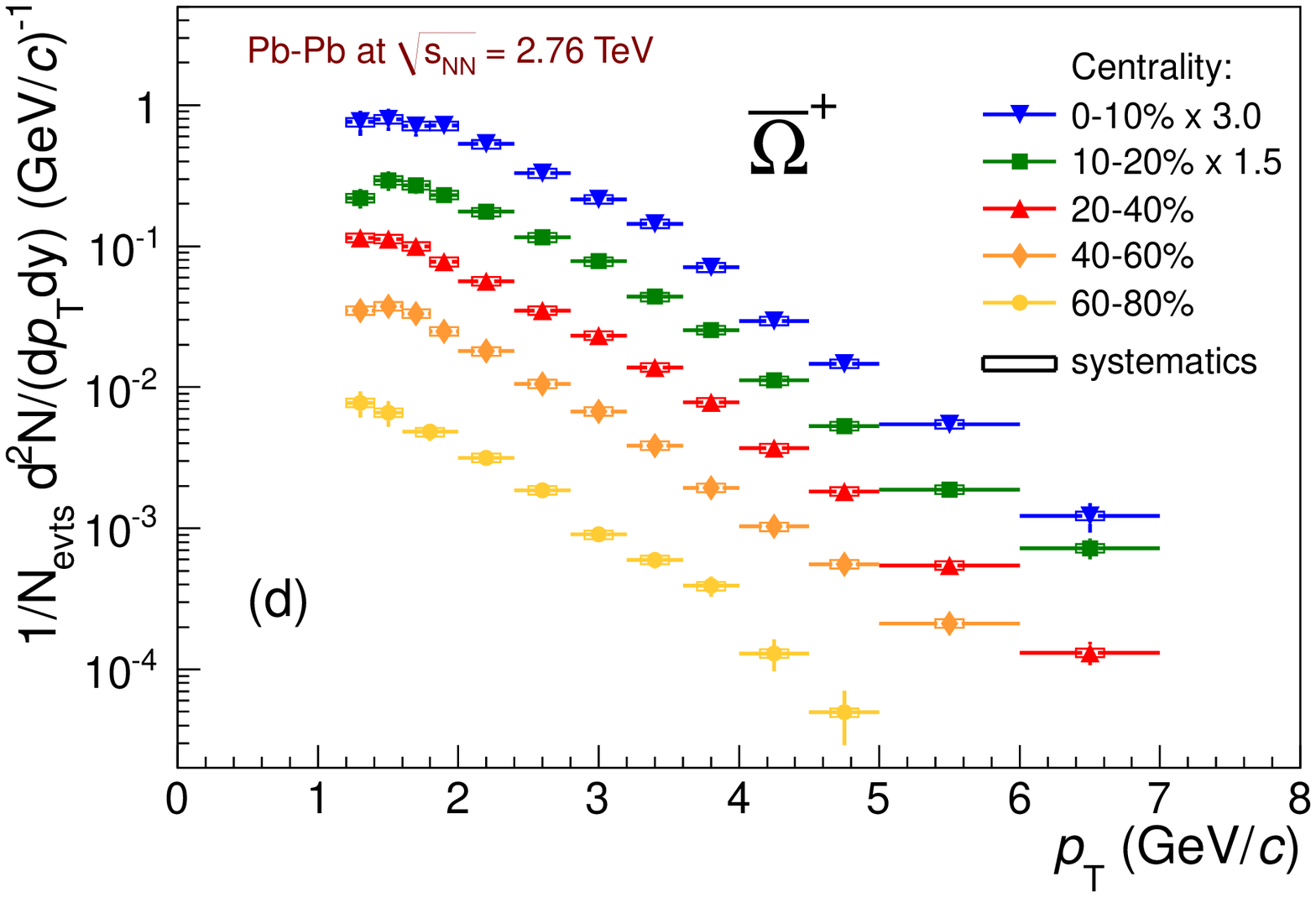}
\caption{\label{Fig_msPbPb_corryieldsvsptcent}
Transverse momentum spectra for \csim\ and \omm\ (a,b) and their anti-particles (c,d) in five diffe\-rent centrality classes,
from the most central (0-10\%) to the most peripheral (60-80\%) \PbPb\ collisions at \mbox{\snn\ = 2.76 TeV},
for $|y|$ $<$ 0.5 at \pt\ $>$ 1.8 \GeV\ and $|y|$ $<$ 0.3 at \pt\ $<$ 1.8 \GeV. 
The statistical error bars are smaller than the symbols for most data points, while the systematic
uncertainties are represented by the open boxes.}
\end{center}
\end{figure}

The corrected \pt\ spectra for each particle species were obtained by dividing bin-by-bin the
raw yield distributions by the \mbox{acceptance $\times$ efficiency} factors determined as described above.
They are shown in Fig.~\ref{Fig_msPbPb_corryieldsvsptcent} for \csim, \csip, \omm\ and \omp,
in the five centrality classes from the most central (0-10\%) to the most peripheral (60-80\%) \PbPb\ collisions.
The values at low \pt\ (below 1.8 \GeV) have been normalized to $|y|$ $<$ 0.5 to make all the points correspond to
a common rapidity window.  
Particle and anti-particle spectra are compatible within errors, as expected at LHC energies.
The \pt\ interval covered in the most central collisions
spans from 0.6 to 8.0 \GeV\ for \csim\ and \csip, and from 1.2 to 7.0 \GeV\ for \omm\ and \omp.
The transverse momentum range of the measurement is limited by the acceptance at low \pt\ and by the available statistics at high \pt.

In order to extract particle yields integrated over the full \pt\ range, the spectra are fitted
using a blast-wave parametrisation~\cite{Schnedermann:1993}. Yields are then calculated by adding to
the integral of the data in the measured \pt\ region, the integral of the fit function outside that region.
The extrapolation to low \pt\ is a much larger fraction of  the yield than that for high \pt:
it contributes between 10-20\% of the final total yields for the \csi, and 35-50\% for \om, depending on centrality.
Other functions of the transverse momentum (exponential, Boltzmann and Tsallis~\cite{Tsallis:1988} parametrisations)
have been used for comparison with the blast-wave shape.
The average difference in the total integrated yield, obtained using the other fit functions, is taken as an estimate
of the systematic uncertainty due to the extrapolation: it is found to be around 7\% for \csi\ and 15\% for \om,
in the worst case of the most peripheral collisions.

The following sources of systematic uncertainty on the final yields have been estimated:
\mbox{i) material} budget in the simulation (4\%),
ii) track selection in the TPC, through the restriction on the number of TPC pad plane clusters
used in the particle reconstruction (1\% for \csi\ and 3\% for \om),
iii) topological and kinematic selection cuts (1\% for \csi\ and 3\% for \om),
iv) for the \om, removal of candidates satisfying the \csi\ mass hypothesis (1\%),
v) signal extraction procedure (1\%),
vi) use of FLUKA~\cite{Ferrari:2005} to correct~\cite{Aamodt:2010bis} the anti-proton absorption cross section in GEANT3~\cite{Brun:1994} (1\%),
vii) centrality dependence of the correction (3\%).
The last contribution is related to the fact that the particle distributions in a given centrality class are
different in the injected Monte Carlo simulations and in the data. 
The total systematic uncertainty, obtained by adding the sources above in quadrature, is 5\% for \csi\ and 7\% for \om, independent of
the \pt\ bin and centrality interval. It has been added in quadrature to the statistical error for each spectra data point before
fitting the distribution and extracting the yields.
An additional systematic error of 7\% (15\%) has been added to the final \csi\ (\om) yield
to take into account the uncertainty due to the extrapolation at low \pt, as mentioned above.

\section{Results and discussion}

The total integrated yields for \csim, \csip, \csim+ \csip, \omm, \omp\ and \omm+ \omp\ 
have been determined in each centrality class, and
are presented in Table~\ref{Table_yields}. 
Statistical and systematic uncertainties are quoted. The systematic errors
include both the contribution due to the correction factors 
and that from the extrapolation to the unmeasured \pt\ region.
Particle and anti-particle yields are found to be compatible within the errors.

\begin{table}[tbh!f]
\centering
\caption{\label{Table_yields}
Total integrated mid-rapidity yields, \dndy, for multi-strange baryons in \PbPb\ collisions at \snn\ = 2.76 TeV,
for different centrality intervals. Both statistical (first) and systematic (second) errors are shown.
For each centrality interval the average number of participants, \avNpart, is also reported~\cite{PbPb276TeVcentrality}.}
\begin{tabular}{lccccc}
\hline\hline
Centrality & {\hspace{4mm} 0-10\% \hspace{4mm}} & {\hspace{4mm} 10-20\% \hspace{4mm}} & {\hspace{4mm} 20-40\% \hspace{4mm}}
           & {\hspace{4mm} 40-60\% \hspace{4mm}} & {\hspace{4mm} 60-80\% \hspace{4mm}} \\
\avNpart   & {356.1 $\pm$ 3.6} & {260.1 $\pm$ 3.9} & {157.2 $\pm$ 3.1} & {68.6 $\pm$ 2.0} & {22.5 $\pm$ 0.8} \\
\hline
\csim      & {\scriptsize 3.34 $\pm$ 0.06 $\pm$ 0.24} & {\scriptsize 2.53 $\pm$ 0.04 $\pm$ 0.18} & {\scriptsize 1.49 $\pm$ 0.02 $\pm$ 0.11} & {\scriptsize 0.53  $\pm$ 0.01  $\pm$ 0.04}  & {\scriptsize 0.124 $\pm$ 0.003 $\pm$ 0.009} \\
\csip      & {\scriptsize 3.28 $\pm$ 0.06 $\pm$ 0.23} & {\scriptsize 2.51 $\pm$ 0.05 $\pm$ 0.18} & {\scriptsize 1.53 $\pm$ 0.02 $\pm$ 0.11} & {\scriptsize 0.54  $\pm$ 0.01  $\pm$ 0.04}  & {\scriptsize 0.120 $\pm$ 0.003 $\pm$ 0.008} \\
\vspace{1mm}
\csim + \csip      & {\scriptsize 6.67 $\pm$ 0.08 $\pm$ 0.47} & {\scriptsize 5.14 $\pm$ 0.06 $\pm$ 0.36} & {\scriptsize 3.03 $\pm$ 0.03 $\pm$ 0.22} & {\scriptsize 1.07  $\pm$ 0.01  $\pm$ 0.08}  & {\scriptsize 0.240 $\pm$ 0.006 $\pm$ 0.019} \\
\omm       & {\scriptsize 0.58 $\pm$ 0.04 $\pm$ 0.09} & {\scriptsize 0.37 $\pm$ 0.03 $\pm$ 0.06} & {\scriptsize 0.23 $\pm$ 0.01 $\pm$ 0.03} & {\scriptsize 0.087 $\pm$ 0.005 $\pm$ 0.014} & {\scriptsize 0.015 $\pm$ 0.002 $\pm$ 0.003} \\
\omp       & {\scriptsize 0.60 $\pm$ 0.05 $\pm$ 0.09} & {\scriptsize 0.40 $\pm$ 0.03 $\pm$ 0.06} & {\scriptsize 0.25 $\pm$ 0.01 $\pm$ 0.03} & {\scriptsize 0.082 $\pm$ 0.005 $\pm$ 0.013} & {\scriptsize 0.017 $\pm$ 0.002 $\pm$ 0.003} \\
\omm + \omp       & {\scriptsize 1.19 $\pm$ 0.06 $\pm$ 0.19} & {\scriptsize 0.78 $\pm$ 0.04 $\pm$ 0.15} & {\scriptsize 0.48 $\pm$ 0.02 $\pm$ 0.08} & {\scriptsize 0.170 $\pm$ 0.007 $\pm$ 0.029} & {\scriptsize 0.032 $\pm$ 0.003 $\pm$ 0.005} \\
\hline\hline
\end{tabular}
\end{table}

The \csi\ and \om\ \pt\ spectra are compared to hydrodynamic model calculations.
The purpose of this comparison is to test the ability of the models to reproduce yields, spectral shape and centrality dependence.
Four models are considered. VISH2+1~\cite{Shen:2011} is a viscous hydrodynamic model, while HKM~\cite{Karpenko:2011,Karpenko:2012}
is an ideal hydrodynamic model similar to VISH2+1 which, in addition, introduces a hadronic cascade (UrQMD~\cite{Bass:1998,Bleicher:1999}) 
following the partonic hydrodynamic phase. 
The Krak{\'o}w model~\cite{Bozek:2012,Bozek:2012bis}, on the other hand, introduces non-equilibrium corrections due to viscosity 
in the transition from a hydrodynamic description to one involving the final state particles. 
EPOS (2.17v3) ~\cite{Werner:2012,Werner:2012bis,Werner:2012ter} aims to be a comprehensive model and event generator, describing all \pt\ domains
with the same dynamical picture: in particular, it incorporates hydrodynamics and models the interaction between
high \pt\ hadrons and the expanding fluid, then uses UrQMD as hadronic cascade model.

\begin{figure}[hbt!f]
\begin{center}
\includegraphics[width=7.8cm]{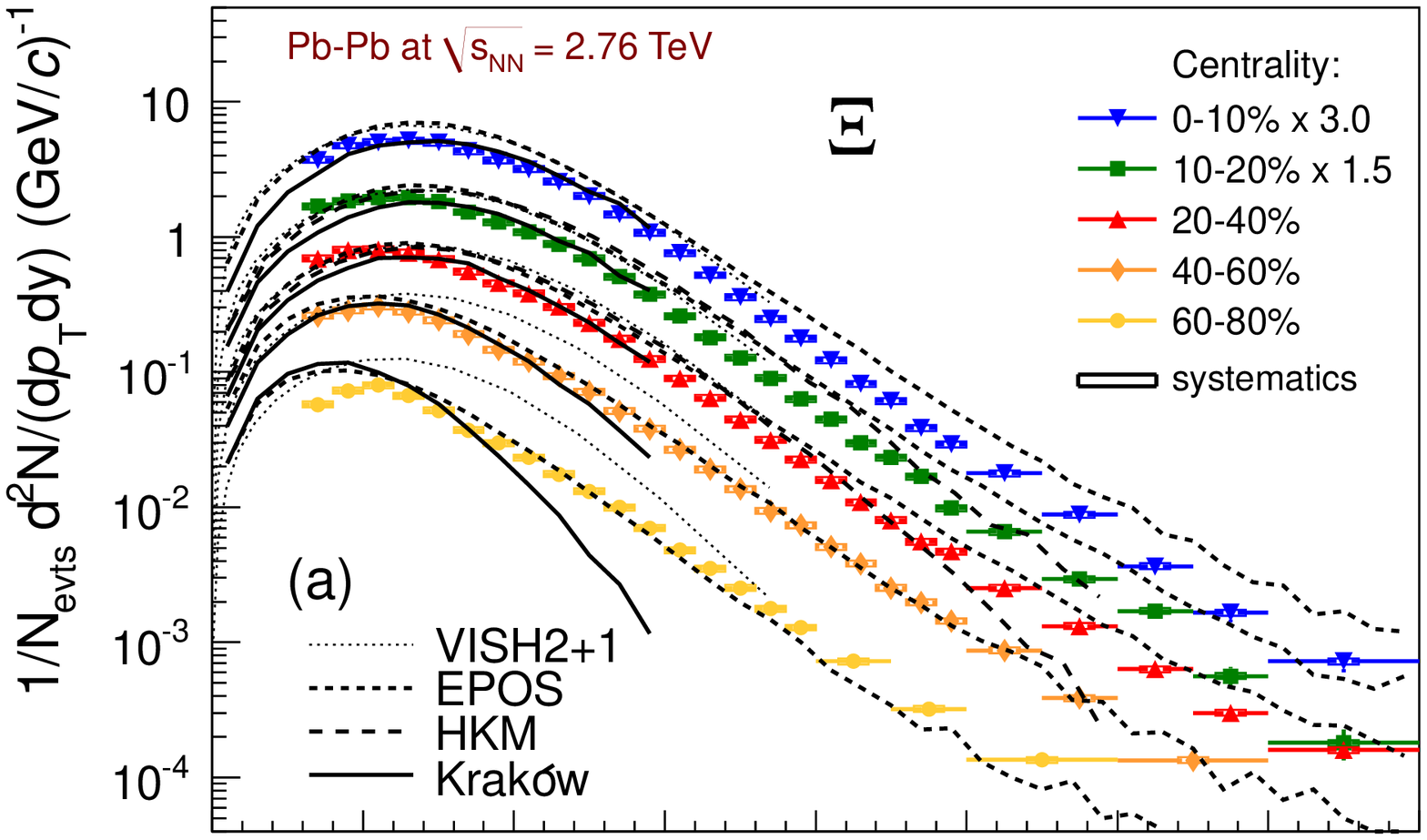}
\includegraphics[width=7.8cm]{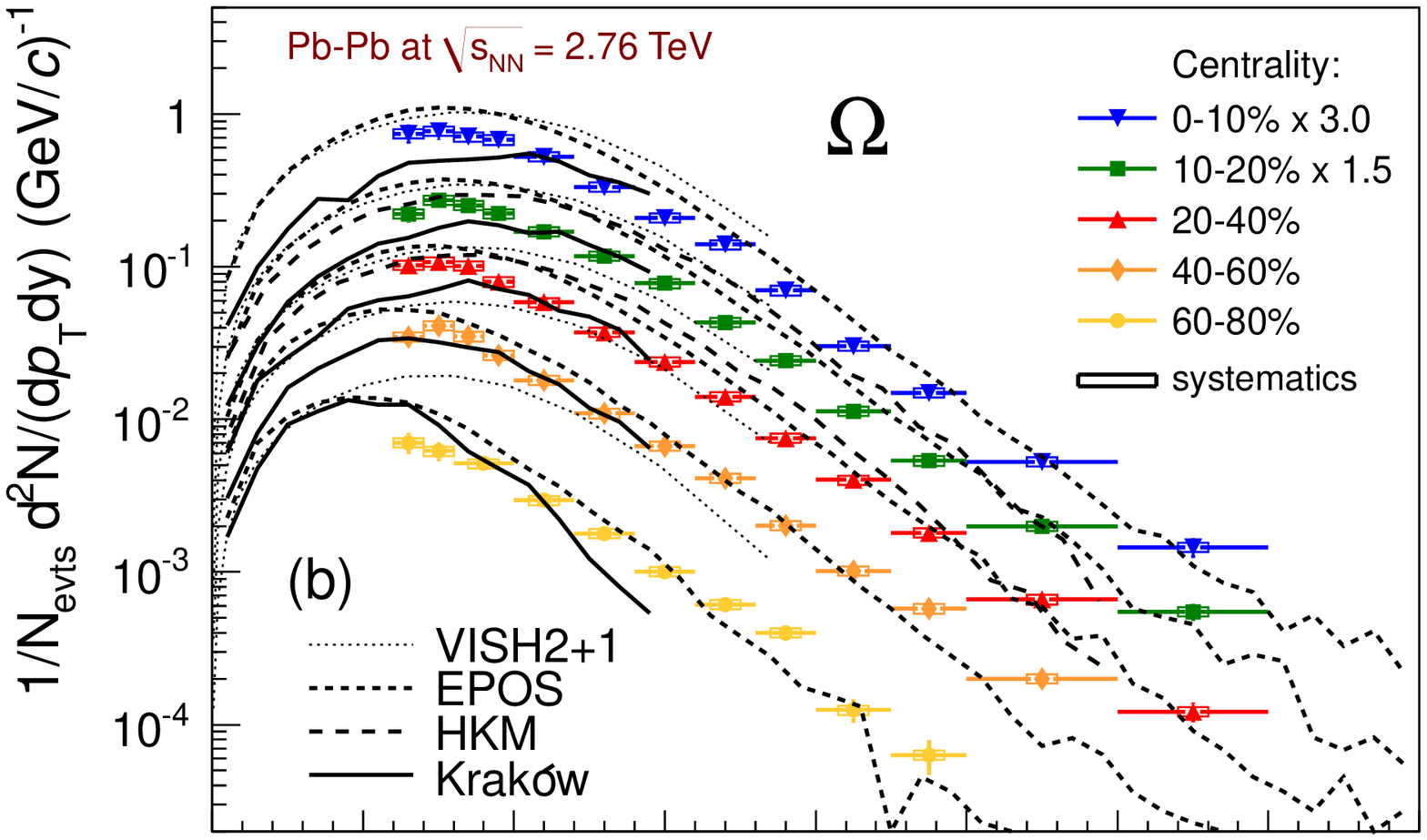}
\includegraphics[width=7.8cm]{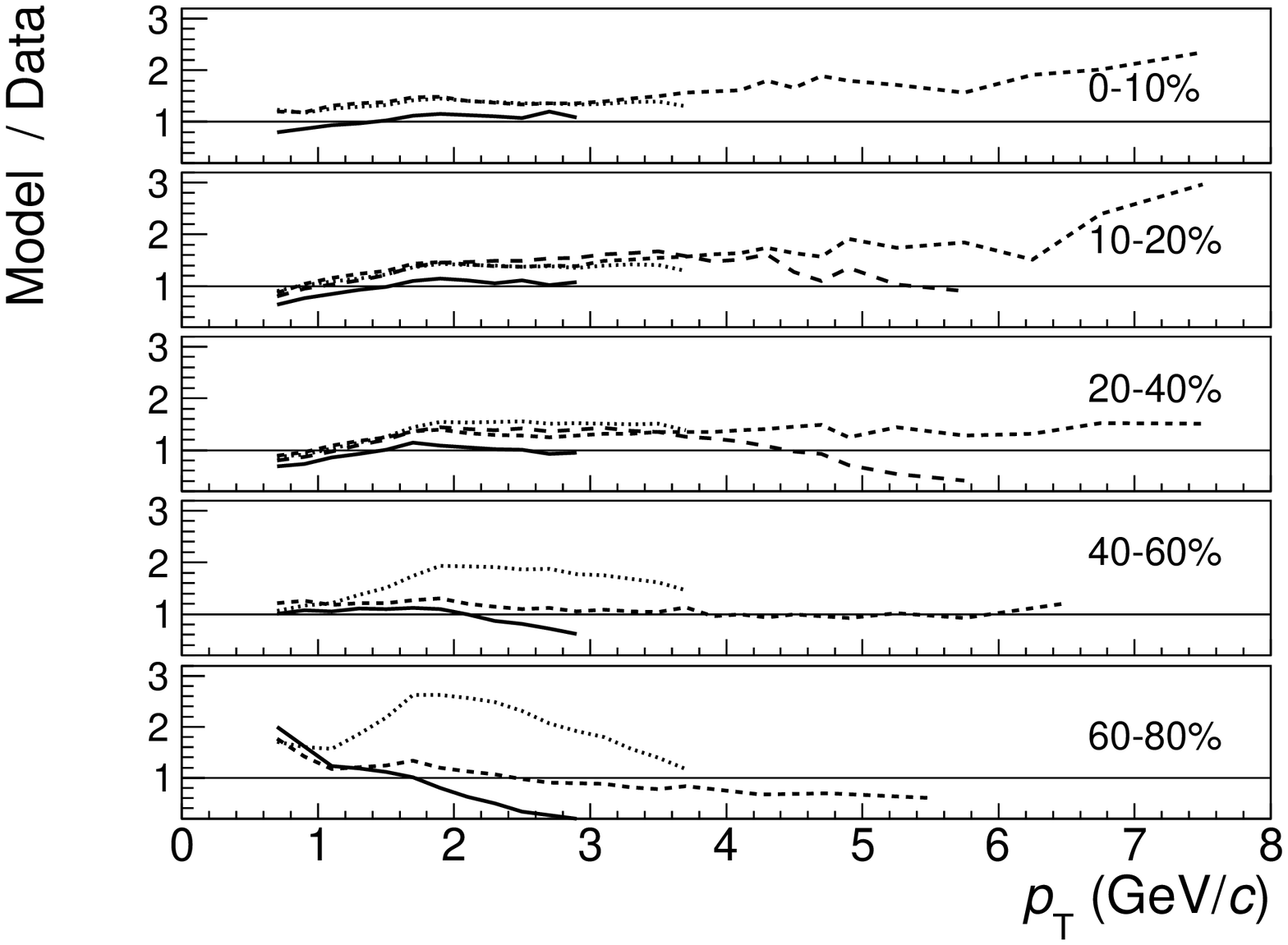}
\includegraphics[width=7.8cm]{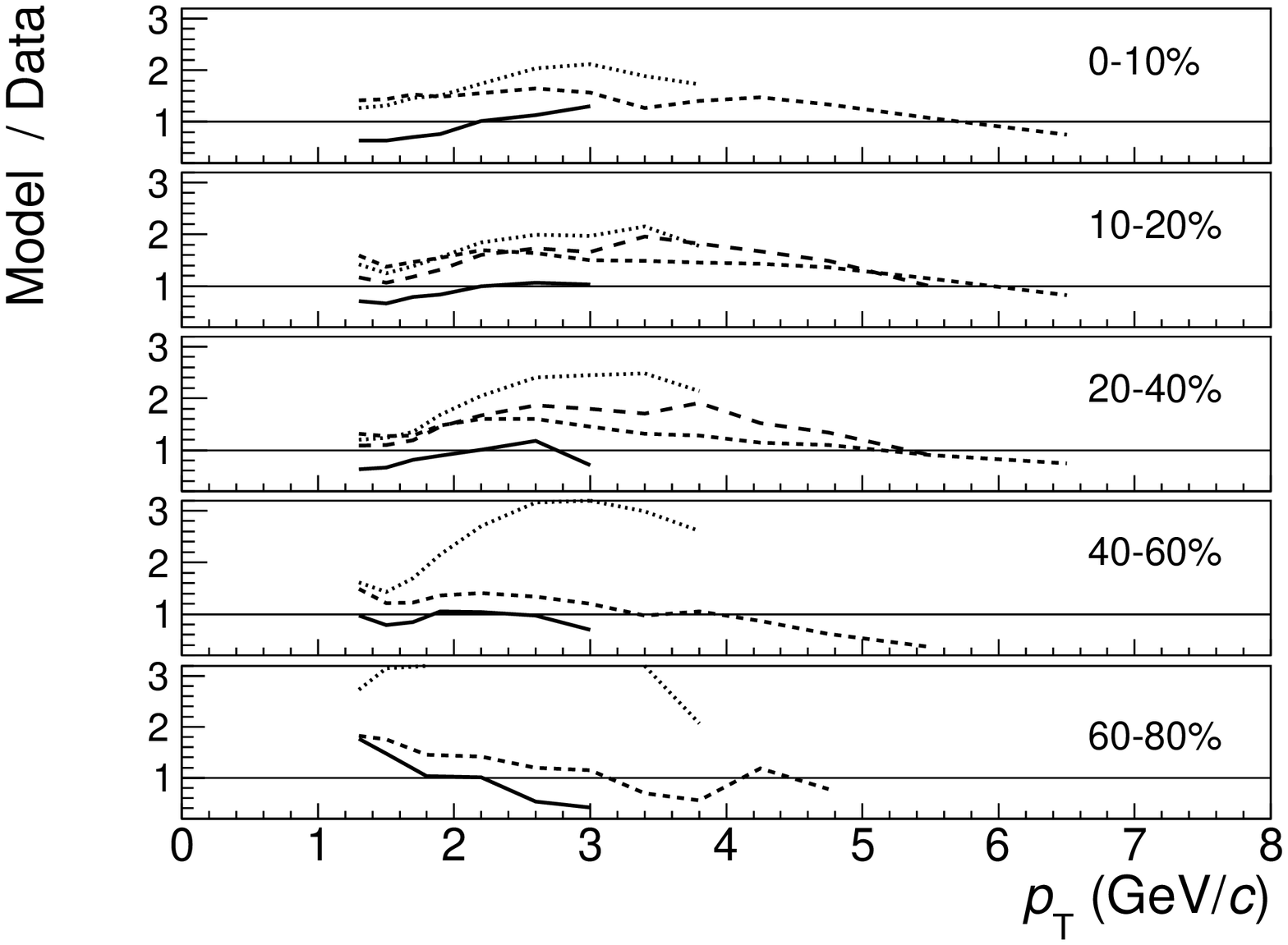}
\caption{\label{Fig_msPbPb_hydrocomp5bins}
Transverse momentum spectra for \csi\ (a) and \om\ (b) hyperons (average of particle and anti-particle)
in five different centrality classes, compared to hydrodynamic models. Ratios of models to data are also shown.}
\end{center}
\end{figure}

The results are shown in Fig.~\ref{Fig_msPbPb_hydrocomp5bins} for \csi\ and \om\ hyperons in different ranges of centrality.
Predictions in each of the data centrality intervals are available for all the models, except for HKM, which is available only 
for the 10-20\% and 20-40\% most central collisions. Moreover, for EPOS the curves correspond to
the average of particle and anti-particle as for the data points, while for the other models only the predictions for
the \csim\ and \omm are available at the time of writing.

We first focus on the most central events (0-10\%). Here, all the available models succeed in describing the shape of
the \csi\ spectrum quite well in the \pt\ range up to 3 \GeV, although only the Krak{\'o}w model correctly reproduces the yield.
This supports the hydrodynamic interpretation of the \pt\ spectra in central collisions at the LHC, which was already
shown to be successful in describing pion, kaon and proton spectra~\cite{Abelev:2013}. 
The description is less successful with the \om. VISH2+1 and EPOS both overestimate the yield, though EPOS reproduces 
the shape; Krak{\'o}w underestimates the yield and does not reproduce the slope.
As we move progressively to less central events, the quality of the agreement remains similar for the \csi, but 
deteriorates for the \om. 
For the \csi, the Krak{\'o}w model describes both the yield and the shape to within about 30\% 
over the centrality range 0-60\%, while it fails to describe the spectrum in the most peripheral class. 
EPOS describes the shape correctly for all centralities and also reproduces the yield in the most peripheral class,
while the other two models give a worse description. 
For the \om, the EPOS and Krak{\'o}w models again provide the most successful description,
reproducing the shape rather well (i.e. to within $\sim$~30\%) in all the centrality classes,
although EPOS consistently overestimates the yields.
As in the case of the \csi, VISH2+1 and HKM provide a less accurate description of the data, though HKM works better than VISH2+1.
Comparing these models gives an insight into the mechanism at work in hyperon production.
VISH2+1, which results in the least successful description, does not include the hadronic cascade mechanism.
The Krak{\'o}w model
indeed provides a good description for both the yields and shapes in the \pt\ range up to 3 \GeV.  
EPOS, on the other hand, includes all these processes and gives the most successful description overall in a wider \pt\ range. 
In this model the aim is to account in a single approach for bulk matter and jets, and the interaction between the two; 
the flux-tubes produced in the
initial hard scattering either escape the medium and hadronize as jets, or contribute to the 
bulk matter where hydrodynamics becomes important. Good agreement has already been observed between EPOS
and ALICE data for pion, kaon and proton spectra in central and semi-central collisions~\cite{Abelev:2013}; in this study
the agreement is confirmed for the \csi\ and \om\ hyperons, and extended to peripheral events.

The strangeness enhancements are defined as ratios of the strange particle yields measured
in \PbPb\ collisions, normalized to the mean number of participant nucleons \avNpart,
to the corresponding quantities in \pp\ interactions at the same energy.
The \pp\ reference values were obtained by interpolating ALICE data at two energies (\s\ = 0.9 and
7 TeV~\cite{Abelev:2012,Aamodt:2011bis}) for the \csi, and STAR data at 200 GeV~\cite{Abelev:2007} and
ALICE data at 7 TeV for the \om.
For both particles, the energy dependence of the PYTHIA yields\footnote{Perugia 2011 tune 88 S350~\cite{Skands:2010} has been used.}
is assumed. Although PYTHIA underestimates the overall yields~\cite{Abelev:2012,Maire:2011}, its energy
dependence is found to be $s^{0.13}$ (which is slightly higher than $s^{0.11}$, obtained for the charged-particle
pseudorapidity density~\cite{Aamodt:2010}): the same power law describes the measured yields
and is therefore used for interpolation.

\begin{figure}[hbt!f]
\begin{center}
\includegraphics[width=10.825cm]{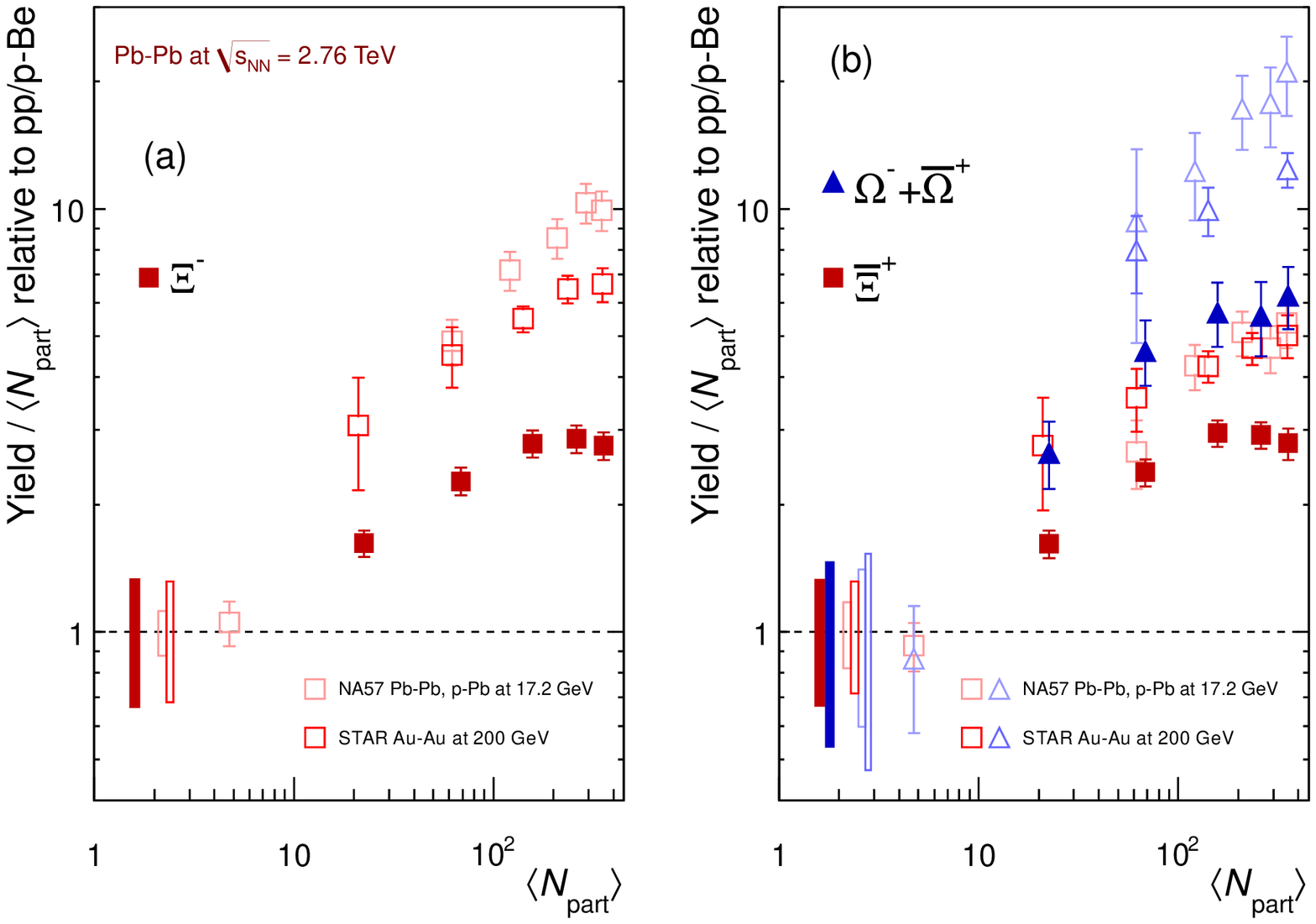}
\hspace{-5mm}
\includegraphics[width=5.3825cm]{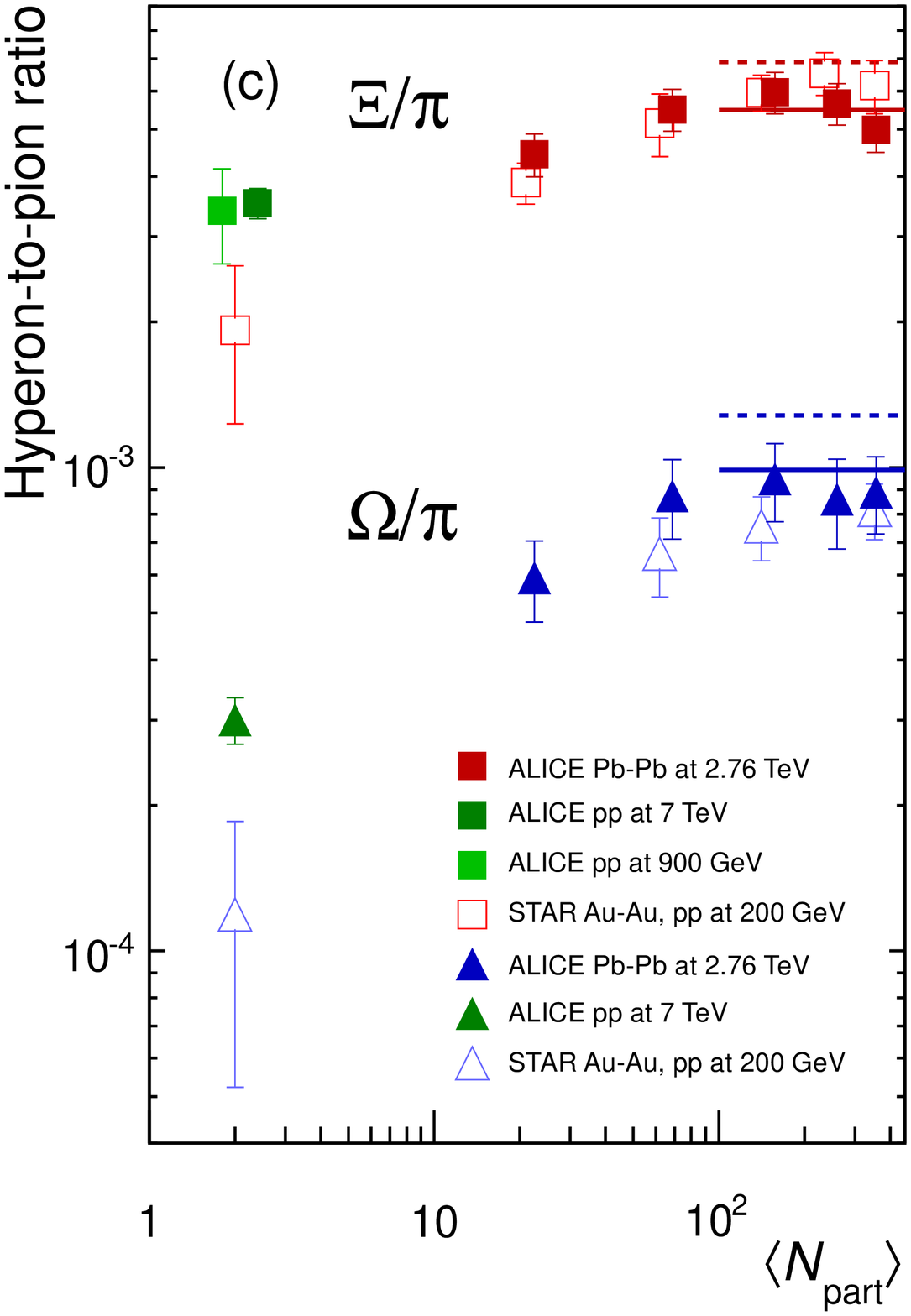}
\vspace{-4mm}
\caption{\label{Fig_msPbPb_enhancementSupp_17_200_2760}
(a,b) Enhancements in the rapidity range $|y|$ $<$ 0.5 as a function of the mean number of participants
\avNpart, showing LHC (ALICE, full symbols), RHIC and SPS (open symbols) data. The LHC data use interpolated
\pp\ values (see text).
Boxes on the dashed line at unity indicate statistical and systematic uncertainties on the \pp\ or \pBe\ reference.
Error bars on the data points represent the corresponding uncertainties for all the heavy-ion measurements 
and those for \pPb\ at the SPS.
(c) Hyperon-to-pion ratios as a function of \avNpart, for \AAcoll\ and \pp\ collisions at LHC and RHIC energies.
The lines mark the thermal model predictions from ~\cite{Andronic:2009} (full line) and ~\cite{Cleymans:2006} (dashed line).}
\end{center}
\end{figure}

Figure~\ref{Fig_msPbPb_enhancementSupp_17_200_2760}a and b show the enhancements for \csim, \csip\ and \omm+ \omp\ in
\PbPb\ collisions at \snn\ = 2.76 TeV (full symbols), as a function of the mean number of participants.
For the \om, particle and anti-particle have been added for the sake of comparison with the
corresponding results at lower energy.
The enhancements are larger than unity for all the particles.
They increase with the strangeness content of the particle, showing the hierarchy already observed at lower energies
and also consistent with the picture of enhanced \ssbar\ pair production in a hot and dense partonic medium.
In addition, the same shape and scale are observed for baryons and anti-baryons (shown for \csim\ and \csip\ in
Fig.~\ref{Fig_msPbPb_enhancementSupp_17_200_2760}), as expected because of the vanishing net-baryon number at the LHC energy.
The centrality dependence shows that the multi-strange particle yields grow faster than linearly with \avNpart,
at least up to the three most central classes (\mbox{\Npart\ $>$ 100-150}), where there are indications of a possible saturation
of the enhancements.
Comparing the ALICE measurements with those from the experiments NA57 at the SPS (\PbPb\ collisions at \mbox{\snn\ = 17.2 GeV})
and STAR at RHIC (\AuAu\ collisions at \snn\ = 200 GeV), represented by the open symbols in 
Fig.~\ref{Fig_msPbPb_enhancementSupp_17_200_2760}a and b, the enhancements are found to decrease with increasing centre-of-mass 
energy, continuing the trend established at lower energies~\cite{Antinori:2006,Antinori:2010,Abelev:2008}. 

The hyperon-to-pion ratios
\mbox{\xitopion\ $\equiv$ (\csim+\csip)/(\pim+\pip)} and
\mbox{\omtopion\ $\equiv$ (\omm+\omp)/(\pim+\pip)},
for \AAcoll\ and \pp\ collisions
both at LHC~\cite{Abelev:2012,Abelev:2013,Aamodt:2011bis,pp7TeVpions:2012,Aamodt:2011ter} 
and RHIC~\cite{Abelev:2007,Abelev:2009bis,Adams:2007} energies, 
are shown in Fig.~\ref{Fig_msPbPb_enhancementSupp_17_200_2760}c
as a function of \avNpart. 
They indicate that different mechanisms contribute to the evolution with centrality of the enhancements as defined above. 
Indeed, the relative production of strangeness in \pp\ collisions is larger than at lower energies. 
The increase in the hyperon-to-pion ratios in \AAcoll\ relative to \pp\ ($\sim$ 1.6 and 3.3 for \csi\ and \om, respectively) 
is about half that of the standard enhancement ratio as defined above. It displays a clear increase in strangeness production relative to \pp, 
rising with centrality up to about \mbox{\avNpart\ $\sim$ 150}, and apparently saturating thereafter.
A small drop is observed in the \xitopion\ ratio for the most central collisions, which is however 
of limited significance given the size of the systematic errors.
Also shown are the predictions for the hyperon-to-pion ratios at the LHC from the thermal models,
based on a grand canonical approach, described 
in ~\cite{Andronic:2009} (full line, with a chemical freeze-out temperature parameter $T$~=~164 MeV) 
and ~\cite{Cleymans:2006} (dashed line, with $T$ = 170 MeV). 
We note that the predictions for $T$ = 164 MeV agree with the present data while, for this temperature, the proton-to-pion 
ratio is overpredicted by about 50\%~\cite{Abelev:2013}. It is now an interesting question whether a grand-canonical
thermal model can give a good description of the complete set of hadron yields in \PbPb\ collisions at LHC energy
with a somewhat lower $T$ value. Alternatively, the low \ptopion\ ratio has been addressed in three \mbox{different} approaches:
\mbox{i) suppression} governed by light quark fugacity in a non-equilibrium \mbox{model~\cite{Rafelski:2013,Rafelski:2013bis},}
\mbox{ii) baryon-antibaryon} annihilation in the hadronic phase, which would have
a stronger effect on protons than on multi-strange particles~\cite{Pan:2012,Becattini:2012bis,Becattini:2012,Steinheimer:2013},
iii) effects due to pre-hadronic flavour-dependent bound states above the QCD transition temperature~\cite{Ratti:2012,Bellwied:2012}.    

\section{Conclusions}

In summary, the measurement of multi-strange baryon production in heavy-ion collisions
at the LHC and the corresponding strangeness enhancements with respect to \pp\ have been presented.
Transverse momentum spectra of mid-rapidity \csim, \csip, \omm\ and \omp\ particles
in \PbPb\ collisions at \snn\ = 2.76~TeV have been measured in five centrality intervals.
The spectra are compared with the predictions from several hydrodynamic models.
It is found that the best agreements are obtained with the Krak{\'o}w and EPOS models, with the 
latter covering a wider \pt\ range.
The yields have been measured to be larger than at RHIC while the hyperon-to-pion ratios
are similar at the two energies, rising with centrality and showing a saturation at \avNpart\ $\sim$ 150.
The values of those ratios for central collisions are found compatible with recent predictions 
from thermal models. 
The enhancements relative to \pp\ increase both with the strangeness content of the baryon and 
with centrality, but are less pronounced than at lower energies.  

%

\newenvironment{acknowledgement}{\relax}{\relax}
\begin{acknowledgement}
\ifpreprint
\section*{Acknowledgements}
\else
\section*{}
\fi
\iffull
  The ALICE collaboration would like to thank all its engineers and technicians for their invaluable contributions to the construction of the experiment and the CERN accelerator teams for the outstanding performance of the LHC complex.
\\
The ALICE collaboration acknowledges the following funding agencies for their support in building and
running the ALICE detector:
 \\
State Committee of Science,  World Federation of Scientists (WFS)
and Swiss Fonds Kidagan, Armenia,
 \\
Conselho Nacional de Desenvolvimento Cient\'{\i}fico e Tecnol\'{o}gico (CNPq), Financiadora de Estudos e Projetos (FINEP),
Funda\c{c}\~{a}o de Amparo \`{a} Pesquisa do Estado de S\~{a}o Paulo (FAPESP);
 \\
National Natural Science Foundation of China (NSFC), the Chinese Ministry of Education (CMOE)
and the Ministry of Science and Technology of China (MSTC);
 \\
Ministry of Education and Youth of the Czech Republic;
 \\
Danish Natural Science Research Council, the Carlsberg Foundation and the Danish National Research Foundation;
 \\
The European Research Council under the European Community's Seventh Framework Programme;
 \\
Helsinki Institute of Physics and the Academy of Finland;
 \\
French CNRS-IN2P3, the `Region Pays de Loire', `Region Alsace', `Region Auvergne' and CEA, France;
 \\
German BMBF and the Helmholtz Association;
\\
General Secretariat for Research and Technology, Ministry of
Development, Greece;
\\
Hungarian OTKA and National Office for Research and Technology (NKTH);
 \\
Department of Atomic Energy and Department of Science and Technology of the Government of India;
 \\
Istituto Nazionale di Fisica Nucleare (INFN) and Centro Fermi -
Museo Storico della Fisica e Centro Studi e Ricerche "Enrico
Fermi", Italy;
 \\
MEXT Grant-in-Aid for Specially Promoted Research, Ja\-pan;
 \\
Joint Institute for Nuclear Research, Dubna;
 \\
National Research Foundation of Korea (NRF);
 \\
CONACYT, DGAPA, M\'{e}xico, ALFA-EC and the EPLANET Program
(European Particle Physics Latin American Network)
 \\
Stichting voor Fundamenteel Onderzoek der Materie (FOM) and the Nederlandse Organisatie voor Wetenschappelijk Onderzoek (NWO), Netherlands;
 \\
Research Council of Norway (NFR);
 \\
Polish Ministry of Science and Higher Education;
 \\
National Authority for Scientific Research - NASR (Autoritatea Na\c{t}ional\u{a} pentru Cercetare \c{S}tiin\c{t}ific\u{a} - ANCS);
 \\
Ministry of Education and Science of Russian Federation, Russian
Academy of Sciences, Russian Federal Agency of Atomic Energy,
Russian Federal Agency for Science and Innovations and The Russian
Foundation for Basic Research;
 \\
Ministry of Education of Slovakia;
 \\
Department of Science and Technology, South Africa;
 \\
CIEMAT, EELA, Ministerio de Econom\'{i}a y Competitividad (MINECO) of Spain, Xunta de Galicia (Conseller\'{\i}a de Educaci\'{o}n),
CEA\-DEN, Cubaenerg\'{\i}a, Cuba, and IAEA (International Atomic Energy Agency);
 \\
Swedish Research Council (VR) and Knut $\&$ Alice Wallenberg
Foundation (KAW);
 \\
Ukraine Ministry of Education and Science;
 \\
United Kingdom Science and Technology Facilities Council (STFC);
 \\
The United States Department of Energy, the United States National
Science Foundation, the State of Texas, and the State of Ohio.
\fi
\end{acknowledgement}

\bibliographystyle{bibstyle}
\bibliography{multistrangePbPbPreprint2}{}

%
%

%

\ifpreprint
\newpage
\appendix
\section{The ALICE Collaboration}
\label{app:collab}
\iffull
  


\begingroup
\small
\begin{flushleft}
B.~Abelev\Irefn{org69}\And
J.~Adam\Irefn{org36}\And
D.~Adamov\'{a}\Irefn{org77}\And
A.M.~Adare\Irefn{org125}\And
M.M.~Aggarwal\Irefn{org81}\And
G.~Aglieri~Rinella\Irefn{org33}\And
M.~Agnello\Irefn{org87}\textsuperscript{,}\Irefn{org104}\And
A.G.~Agocs\Irefn{org124}\And
A.~Agostinelli\Irefn{org25}\And
Z.~Ahammed\Irefn{org120}
% ...
\renewcommand\labelenumi{\textsuperscript{\theenumi}~}

\section*{Affiliation notes}
\renewcommand\theenumi{\roman{enumi}}
\begin{Authlist}
\item \Adef{0}Deceased
\item \Adef{idp3706176}{Also at: M.V.Lomonosov Moscow State University, D.V.Skobeltsyn Institute of Nuclear Physics, Moscow, Russia}
\item \Adef{idp3951424}{Also at: University of Belgrade, Faculty of Physics and "Vin\v{c}a" Institute of Nuclear Sciences, Belgrade, Serbia}
\item \Adef{idp4240112}{Permanent address: Konkuk University, Seoul, Korea}
\item \Adef{idp4765344}{Also at: Institute of Theoretical Physics, University of Wroclaw, Wroclaw, Poland}
\end{Authlist}

\section*{Collaboration Institutes}
\renewcommand\theenumi{\arabic{enumi}~}
\begin{Authlist}

\item \Idef{org1}A. I. Alikhanyan National Science Laboratory (Yerevan Physics Institute) Foundation, Yerevan, Armenia
\item \Idef{org2}Benem\'{e}rita Universidad Aut\'{o}noma de Puebla, Puebla, Mexico
\item \Idef{org3}Bogolyubov Institute for Theoretical Physics, Kiev, Ukraine
\item \Idef{org4}Budker Institute for Nuclear Physics, Novosibirsk, Russia
\item \Idef{org5}California Polytechnic State University, San Luis Obispo, California, United States
\item \Idef{org6}Central China Normal University, Wuhan, China
\item \Idef{org7}Centre de Calcul de l'IN2P3, Villeurbanne, France 
\item \Idef{org8}Centro de Aplicaciones Tecnol\'{o}gicas y Desarrollo Nuclear (CEADEN), Havana, Cuba
\item \Idef{org9}Centro de Investigaciones Energ\'{e}ticas Medioambientales y Tecnol\'{o}gicas (CIEMAT), Madrid, Spain
\item \Idef{org10}Centro de Investigaci\'{o}n y de Estudios Avanzados (CINVESTAV), Mexico City and M\'{e}rida, Mexico
\item \Idef{org11}Centro Fermi - Museo Storico della Fisica e Centro Studi e Ricerche ``Enrico Fermi'', Rome, Italy
\item \Idef{org12}Chicago State University, Chicago, United States
\item \Idef{org13}Commissariat \`{a} l'Energie Atomique, IRFU, Saclay, France
\item \Idef{org14}COMSATS Institute of Information Technology (CIIT), Islamabad, Pakistan
\item \Idef{org15}Departamento de F\'{\i}sica de Part\'{\i}culas and IGFAE, Universidad de Santiago de Compostela, Santiago de Compostela, Spain
\item \Idef{org16}Department of Physics Aligarh Muslim University, Aligarh, India
\item \Idef{org17}Department of Physics and Technology, University of Bergen, Bergen, Norway
\item \Idef{org18}Department of Physics, Ohio State University, Columbus, Ohio, United States
\item \Idef{org19}Department of Physics, Sejong University, Seoul, South Korea
\item \Idef{org20}Department of Physics, University of Oslo, Oslo, Norway
\item \Idef{org21}Dipartimento di Fisica dell'Universit\`{a} and Sezione INFN, Cagliari, Italy
\item \Idef{org22}Dipartimento di Fisica dell'Universit\`{a} and Sezione INFN, Trieste, Italy
\item \Idef{org23}Dipartimento di Fisica dell'Universit\`{a} and Sezione INFN, Turin, Italy
\item \Idef{org24}Dipartimento di Fisica dell'Universit\`{a} `La Sapienza` and Sezione INFN, Rome, Italy
\item \Idef{org25}Dipartimento di Fisica e Astronomia dell'Universit\`{a} and Sezione INFN, Bologna, Italy
\item \Idef{org26}Dipartimento di Fisica e Astronomia dell'Universit\`{a} and Sezione INFN, Catania, Italy
\item \Idef{org27}Dipartimento di Fisica e Astronomia dell'Universit\`{a} and Sezione INFN, Padova, Italy
\item \Idef{org28}Dipartimento di Fisica `E.R.~Caianiello' dell'Universit\`{a} and Gruppo Collegato INFN, Salerno, Italy
\item \Idef{org29}Dipartimento di Scienze e Innovazione Tecnologica dell'Universit\`{a} del Piemonte Orientale and Gruppo Collegato INFN, Alessandria, Italy
\item \Idef{org30}Dipartimento Interateneo di Fisica `M.~Merlin' and Sezione INFN, Bari, Italy
\item \Idef{org31}Division of Experimental High Energy Physics, University of Lund, Lund, Sweden
\item \Idef{org32}Eberhard Karls Universit\"{a}t T\"{u}bingen, T\"{u}bingen, Germany
\item \Idef{org33}European Organization for Nuclear Research (CERN), Geneva, Switzerland
\item \Idef{org34}Faculty of Engineering, Bergen University College, Bergen, Norway
\item \Idef{org35}Faculty of Mathematics, Physics and Informatics, Comenius University, Bratislava, Slovakia
\item \Idef{org36}Faculty of Nuclear Sciences and Physical Engineering, Czech Technical University in Prague, Prague, Czech Republic
\item \Idef{org37}Faculty of Science, P.J.~\v{S}af\'{a}rik University, Ko\v{s}ice, Slovakia
\item \Idef{org38}Frankfurt Institute for Advanced Studies, Johann Wolfgang Goethe-Universit\"{a}t Frankfurt, Frankfurt, Germany
\item \Idef{org39}Gangneung-Wonju National University, Gangneung, South Korea
\item \Idef{org40}Helsinki Institute of Physics (HIP), Helsinki, Finland
\item \Idef{org41}Hiroshima University, Hiroshima, Japan
\item \Idef{org42}Indian Institute of Technology Bombay (IIT), Mumbai, India
\item \Idef{org43}Indian Institute of Technology Indore, India (IITI)
\item \Idef{org44}Institut de Physique Nucl\'{e}aire d'Orsay (IPNO), Universit\'{e} Paris-Sud, CNRS-IN2P3, Orsay, France
\item \Idef{org45}Institut f\"{u}r Informatik, Johann Wolfgang Goethe-Universit\"{a}t Frankfurt, Frankfurt, Germany
\item \Idef{org46}Institut f\"{u}r Kernphysik, Johann Wolfgang Goethe-Universit\"{a}t Frankfurt, Frankfurt, Germany
\item \Idef{org47}Institut f\"{u}r Kernphysik, Technische Universit\"{a}t Darmstadt, Darmstadt, Germany
\item \Idef{org48}Institut f\"{u}r Kernphysik, Westf\"{a}lische Wilhelms-Universit\"{a}t M\"{u}nster, M\"{u}nster, Germany
\item \Idef{org49}Institut Pluridisciplinaire Hubert Curien (IPHC), Universit\'{e} de Strasbourg, CNRS-IN2P3, Strasbourg, France
\item \Idef{org50}Institute for High Energy Physics, Protvino, Russia
\item \Idef{org51}Institute for Nuclear Research, Academy of Sciences, Moscow, Russia
\item \Idef{org52}Institute for Subatomic Physics of Utrecht University, Utrecht, Netherlands
\item \Idef{org53}Institute for Theoretical and Experimental Physics, Moscow, Russia
\item \Idef{org54}Institute of Experimental Physics, Slovak Academy of Sciences, Ko\v{s}ice, Slovakia
\item \Idef{org55}Institute of Physics, Academy of Sciences of the Czech Republic, Prague, Czech Republic
\item \Idef{org56}Institute of Physics, Bhubaneswar, India
\item \Idef{org57}Institute of Space Science (ISS), Bucharest, Romania
\item \Idef{org58}Instituto de Ciencias Nucleares, Universidad Nacional Aut\'{o}noma de M\'{e}xico, Mexico City, Mexico
\item \Idef{org59}Instituto de F\'{\i}sica, Universidad Nacional Aut\'{o}noma de M\'{e}xico, Mexico City, Mexico
\item \Idef{org60}iThemba LABS, National Research Foundation, Somerset West, South Africa
\item \Idef{org61}Joint Institute for Nuclear Research (JINR), Dubna, Russia
\item \Idef{org62}Korea Institute of Science and Technology Information, Daejeon, South Korea
\item \Idef{org63}KTO Karatay University, Konya, Turkey
\item \Idef{org64}Laboratoire de Physique Corpusculaire (LPC), Clermont Universit\'{e}, Universit\'{e} Blaise Pascal, CNRS--IN2P3, Clermont-Ferrand, France
\item \Idef{org65}Laboratoire de Physique Subatomique et de Cosmologie (LPSC), Universit\'{e} Joseph Fourier, CNRS-IN2P3, Institut Polytechnique de Grenoble, Grenoble, France
\item \Idef{org66}Laboratori Nazionali di Frascati, INFN, Frascati, Italy
\item \Idef{org67}Laboratori Nazionali di Legnaro, INFN, Legnaro, Italy
\item \Idef{org68}Lawrence Berkeley National Laboratory, Berkeley, California, United States
\item \Idef{org69}Lawrence Livermore National Laboratory, Livermore, California, United States
\item \Idef{org70}Moscow Engineering Physics Institute, Moscow, Russia
\item \Idef{org71}National Centre for Nuclear Studies, Warsaw, Poland
\item \Idef{org72}National Institute for Physics and Nuclear Engineering, Bucharest, Romania
\item \Idef{org73}National Institute of Science Education and Research, Bhubaneswar, India
\item \Idef{org74}Niels Bohr Institute, University of Copenhagen, Copenhagen, Denmark
\item \Idef{org75}Nikhef, National Institute for Subatomic Physics, Amsterdam, Netherlands
\item \Idef{org76}Nuclear Physics Group, STFC Daresbury Laboratory, Daresbury, United Kingdom
\item \Idef{org77}Nuclear Physics Institute, Academy of Sciences of the Czech Republic, \v{R}e\v{z} u Prahy, Czech Republic
\item \Idef{org78}Oak Ridge National Laboratory, Oak Ridge, Tennessee, United States
\item \Idef{org79}Petersburg Nuclear Physics Institute, Gatchina, Russia
\item \Idef{org80}Physics Department, Creighton University, Omaha, Nebraska, United States
\item \Idef{org81}Physics Department, Panjab University, Chandigarh, India
\item \Idef{org82}Physics Department, University of Athens, Athens, Greece
\item \Idef{org83}Physics Department, University of Cape Town, Cape Town, South Africa
\item \Idef{org84}Physics Department, University of Jammu, Jammu, India
\item \Idef{org85}Physics Department, University of Rajasthan, Jaipur, India
\item \Idef{org86}Physikalisches Institut, Ruprecht-Karls-Universit\"{a}t Heidelberg, Heidelberg, Germany
\item \Idef{org87}Politecnico di Torino, Turin, Italy
\item \Idef{org88}Purdue University, West Lafayette, Indiana, United States
\item \Idef{org89}Pusan National University, Pusan, South Korea
\item \Idef{org90}Research Division and ExtreMe Matter Institute EMMI, GSI Helmholtzzentrum f\"ur Schwerionenforschung, Darmstadt, Germany
\item \Idef{org91}Rudjer Bo\v{s}kovi\'{c} Institute, Zagreb, Croatia
\item \Idef{org92}Russian Federal Nuclear Center (VNIIEF), Sarov, Russia
\item \Idef{org93}Russian Research Centre Kurchatov Institute, Moscow, Russia
\item \Idef{org94}Saha Institute of Nuclear Physics, Kolkata, India
\item \Idef{org95}School of Physics and Astronomy, University of Birmingham, Birmingham, United Kingdom
\item \Idef{org96}Secci\'{o}n F\'{\i}sica, Departamento de Ciencias, Pontificia Universidad Cat\'{o}lica del Per\'{u}, Lima, Peru
\item \Idef{org97}Sezione INFN, Bari, Italy
\item \Idef{org98}Sezione INFN, Bologna, Italy
\item \Idef{org99}Sezione INFN, Cagliari, Italy
\item \Idef{org100}Sezione INFN, Catania, Italy
\item \Idef{org101}Sezione INFN, Padova, Italy
\item \Idef{org102}Sezione INFN, Rome, Italy
\item \Idef{org103}Sezione INFN, Trieste, Italy
\item \Idef{org104}Sezione INFN, Turin, Italy
\item \Idef{org105}SUBATECH, Ecole des Mines de Nantes, Universit\'{e} de Nantes, CNRS-IN2P3, Nantes, France
\item \Idef{org106}Suranaree University of Technology, Nakhon Ratchasima, Thailand
\item \Idef{org107}Technical University of Split FESB, Split, Croatia
\item \Idef{org108}The Henryk Niewodniczanski Institute of Nuclear Physics, Polish Academy of Sciences, Cracow, Poland
\item \Idef{org109}The University of Texas at Austin, Physics Department, Austin, TX, United States
\item \Idef{org110}Universidad Aut\'{o}noma de Sinaloa, Culiac\'{a}n, Mexico
\item \Idef{org111}Universidade de S\~{a}o Paulo (USP), S\~{a}o Paulo, Brazil
\item \Idef{org112}Universidade Estadual de Campinas (UNICAMP), Campinas, Brazil
\item \Idef{org113}University of Houston, Houston, Texas, United States
\item \Idef{org114}University of Jyv\"{a}skyl\"{a}, Jyv\"{a}skyl\"{a}, Finland
\item \Idef{org115}University of Tennessee, Knoxville, Tennessee, United States
\item \Idef{org116}University of Tokyo, Tokyo, Japan
\item \Idef{org117}University of Tsukuba, Tsukuba, Japan
\item \Idef{org118}Universit\'{e} de Lyon, Universit\'{e} Lyon 1, CNRS/IN2P3, IPN-Lyon, Villeurbanne, France
\item \Idef{org119}V.~Fock Institute for Physics, St. Petersburg State University, St. Petersburg, Russia
\item \Idef{org120}Variable Energy Cyclotron Centre, Kolkata, India
\item \Idef{org121}Vestfold University College, Tonsberg, Norway
\item \Idef{org122}Warsaw University of Technology, Warsaw, Poland
\item \Idef{org123}Wayne State University, Detroit, Michigan, United States
\item \Idef{org124}Wigner Research Centre for Physics, Hungarian Academy of Sciences, Budapest, Hungary
\item \Idef{org125}Yale University, New Haven, Connecticut, United States
\item \Idef{org126}Yonsei University, Seoul, South Korea
\item \Idef{org127}Zentrum f\"{u}r Technologietransfer und Telekommunikation (ZTT), Fachhochschule Worms, Worms, Germany
\end{Authlist}
\endgroup

\fi
\fi

\end{document}